# Quantum Sensing Enhancement through a Nuclear Spin Register in Nitrogen-Vacancy Centers in Diamond

Jonathan Kenny[1], Feifei Zhou[1], Ruihua He[1], Fedor Jelezko*[2], Teck Seng Koh*[1], Weibo Gao*[1,3,4,5]

[1]Division of Physics and Applied Physics, School of Physical and Mathematical Sciences, Nanyang Technological University; Singapore, 637371, Singapore

[2]Institute for Quantum Optics and Centre for Integrated Quantum Science and technology (IQST), Ulm University, Albert-Einstein-Allee 11, 89081 Ulm, Germany

[3]School of Electrical and Electronic Engineering, Nanyang Technological University, Singapore, Singapore

[4]Centre for Quantum Technologies, Nanyang Technological University, Singapore, Singapore

[5]Quantum Science and Engineering Centre (QSec), College of engineering, Nanyang Technological University, Singapore, Singapore

*Corresponding author. Email: fedor.jelezko@uni-ulm.de, kohteckseng@ntu.edu.sg, wbgao@ntu.edu.sg

## Abstract

Quantum sensing has witnessed rapid development and transition from laboratories to practical applications in the past decade. Applications of quantum sensors, ranging from nanotechnologies to biosensing, are expected to benefit from quantum sensors' unprecedented spatial resolution and sensitivity. Solid-state spin systems are particularly attractive platforms for quantum sensing technologies because room temperature operation is viable while reaching the quantum limits of sensitivity. Among various solid-state spin systems, nitrogen-vacancy (NV) centers in diamond demonstrated high-fidelity initialization, coherent control and high contrast readout of the electron spin state. However, electron spin coherence due to noise from the surrounding spin bath and this environment effect limits the sensitivity of NV centers. Thus, a critical task in NV center-based quantum sensing is sensitivity enhancement through coherence protection. Several strategies such as dynamical decoupling techniques, feedback control, and nuclear spin assisted sensing protocols, have been developed and realized for this task. Among these strategies, nuclear spin assisted protocols have demonstrated greater enhancement of electron spin coherence. In addition, the electron and nuclear spin pair of an NV center in diamond naturally allows the application of the nuclear spin assisted sensitivity enhancement protocol. Owing to long nuclear coherence times, further enhancement of sensitivity can be achieved by exploiting active nuclear spins (e.g. $^{14}N$, $^{13}C$) in the proximity of an NV center as memory ancillas when coupled with the NV center. Here, we review the spin properties of NV centers, mechanisms of the nuclear spin assisted protocol and its gate variation, and the status of quantum sensing applications in high-resolution nuclear spin spectroscopy, atomic imaging, magnetic field sensing. We discuss the potential for commercialization, current challenges in sensitivity enhancement and conclude with future research directions for promoting the development of nuclear spin assisted protocol and its integration into industrial applications.

# 1. Introduction

Quantum sensing exploits the high susceptibility of a quantum system to external fields, allowing it to approach fundamental levels of sensitivity. The key figure of merit of quantum sensors is the smallest detectable signal, also known as the sensitivity of the quantum sensor. Many quantum systems have been explored for use as quantum sensors. For example, the trapped ions platform [1, 2] has been widely used in electric field-sensitive detection; the atomic vapor system [3,4,5] can perform magnetic field quantum sensing at attotesla sensitivity; and the superconducting qubits platform [6,7,8] can achieve magnetic field sensitivity at the picotesla level. However, the trapped ions platform requires close surface proximity to perform, which presents great challenges in practical applications. Meanwhile, the trapped ions platform is a single sensor, whose sensitivity is not comparable to ensemble sensors such as atomic vapors. Furthermore, the atomic vapor platform relies on thermal vapors that necessitate operating conditions above room temperature, while the superconducting qubits system needs to be cooled down to sub-millikelvin temperature to ensure its high sensitivity. Therefore, each of the aforementioned platforms comes with its advantages and constraints. In recent years, the platform of solid-state spin sensors based on atomic defects in crystals has garnered significant attention and interest for its high magnetic field sensitivity at the picotesla level [9, 10, 11, 12] and room temperature operation [13].

One of the most representative solid-state spin sensors is the nitrogen-vacancy (NV) center in diamond, which has already been employed in the quantum sensing of magnetic fields [14, 15, 16, 17], electric fields [18, 19, 20], temperature [21, 22, 23], strain [24, 24, 26], and paramagnetic spins [27, 28, 29]. It is a hybrid spin system comprising an electron spin with surrounding nitrogen and carbon nuclear spins. Since it was first proposed in 2008 [9, 30], studies of electron spins in NV centers have demonstrated high-fidelity initialization, readout and coherent control [31, 32, 33, 34, 35]. However, electron spins in NV centers have limited coherence times (longest reported $T_2$ of approximately 7 2.4 ms [36]) at room temperature arising from noise in its internal spin bath and phonon coupling, resulting in reduced sensing performance. Thus, in the field of NV centers, a critical task is to enhance sensitivity through coherence protection. Several strategies have been developed and realized for this task, including dynamical decoupling [37, 38], feedback control [39], quantum heterodyne [40, 41], quantum error correction [42, 43] and nuclear spin assisted sensing protocols [44, 45, 46]. In comparison with the other mentioned strategies, the nuclear spin assisted protocol demonstrates greater enhancement of electron spin coherence time by utilizing nuclear spins as ancilla memory qubits due to their longer coherence times ($T_2 \sim 1$ ms [47]).

This review aims to provide a summary of recent advances in quantum sensing enhancement through the nuclear spin register methods for NV centers in diamond. In section 2, we first review the energy level structure of a single NV center, its initialization, electron spin readout, and nuclear spin coupling and control. Section 3 describes the hyperfine interaction, fundamental mechanisms of the electron-nuclear spin system, followed by the first breakthrough of the nuclear spin assisted protocol in beating the standard quantum limit. Section 4 compares different strategies for nuclear spin assisted quantum sensing enhancement and discusses several breakthrough applications. Finally, we discuss the main challenges and future research endeavors toward quantum sensing enhancement via the nuclear spin assisted protocol.

## 2. Basic concepts of the hybrid spin system

### 2.1. Electron and Nuclear Spin Hamiltonian

The electron ($S$) and nuclear ($I$) spin of a single NV center in a magnetic field ($B$) can be described by the Hamiltonian

$$H = DS_z^2 - QI_z^2 + \gamma_e \vec{B} \cdot \vec{S} - \gamma_N \vec{B} \cdot \vec{I} + \vec{S} \cdot \overleftrightarrow{A} \cdot \vec{I}$$

where the first two terms are the zero-field splitting and quadrupole interaction terms of the electron and nuclear spin, the third and fourth terms are the electron and nuclear spin Zeeman terms, and the last term is the hyperfine interaction between the electron and nuclear spin. Here, $\gamma_e = 2.8\ MHzG^{-1}$ and $\gamma_n = 307.7(-431.7)\ HzG^{-1}$ are the electron and $^{14}N(^{15}N)$ nuclear spin gyromagnetic ratios, $D = 2.87\ GHz$ is the zero-field energy splitting [48] and $Q = 5.04\ MHz$ is the quadrupole energy split. $\overleftrightarrow{A}$ is the hyperfine field tensor, while $\vec{S}, \vec{I}$ are the electron and nuclear spin operators and $S_z, I_z$ are the corresponding $z$-direction operators. In the nuclear spin assisted sensing enhancement protocol, the electron spin $S = 1$, is coupled to $^{15}N$ nuclear spin with spin quantum number $I = \frac{1}{2}$. In some cases, coupling to $^{14}N$ nuclear spin with spin quantum number $I = 1$ is observed.

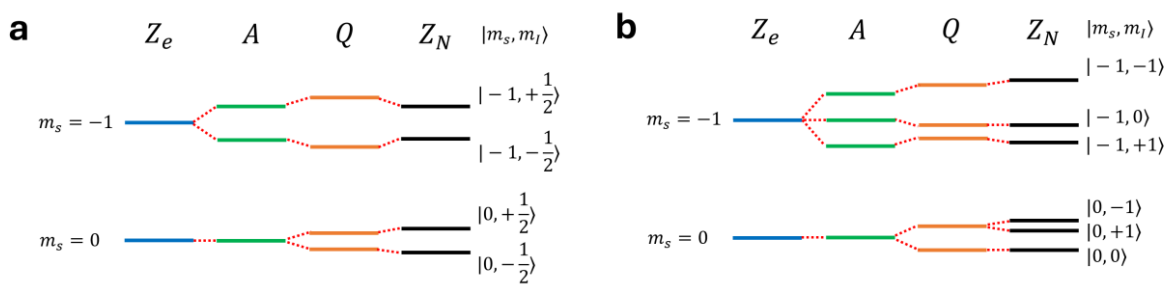

**Figure 1:** Schematic illustration of energy sublevels of a typical NV center with the electron spin $S = 1$ and nuclear spin species a) $^{15}N, I = \frac{1}{2}$ b) $^{14}N, I = 1$. $Z_e$ (blue lines) and $Z_n$ (orange lines) are the energy sublevel splitting due to electron and nuclear Zeeman eefects. $Q$ (green lines) and $A$ (black lines) represent the quadrupole and the hyperfine interaction effect on the energy sublevels, respectively. The electron spin state $m_s = +1$ is not shown. The dotted lines indicate the energy shifts due to the various Hamiltonian terms. States are labeled by $|m_s, m_i\rangle$ which are good quantum numbers since $D \gg A$ in the presence of low external magnetic field.

Figure 1 shows the relevant electron and nuclear spin energy sublevels in an NV center for both nuclear spin number $I = \frac{1}{2}$ and $I = 1$. The application of a magnetic field lifts the degeneracy of the electron and nuclear spin sublevels, labeled as $m_s = 0, -1$ and $m_I = +\frac{1}{2}, -\frac{1}{2}$ for $I = \frac{1}{2}$ and $m_I = 0, \pm 1$ for $I = 1$. In many NV center sensing protocols, either $m_s = +1$ or $m_s = -1$ manifold can be preferentially chosen due to experimental choice. In both $m_s = \pm 1$ manifolds, the energy splitting dictated by the strength of the hyperfine interaction can be observed. Meanwhile, in the $m_s = 0$ manifold, the hyperfine tensor is zero, and thus the energy splitting is dominated by the quadrupole interaction and the Zeeman splitting. Identification of the transition frequencies enables coherent control and population manipulation among these sublevels.

## 2.2. Electron spin readout

Optical readout of an NV center relies on the metastable shelving state $|s\rangle$ of the energy level diagram as shown in Figure 2a. By using a 532 nm laser, the ground state $|g\rangle$ is excited to the excited state $|e\rangle$ in the NV center. This excitation is spin conserving, meaning that the population of $m_s = 0, \pm 1$ is preserved across the ground and excited states. The differing intersystem crossing rates from the excited states $|e\rangle$ to the shelving state $|s\rangle$ results in shorter lifetimes for electrons in $m_s = \pm 1$ states which decay primarily due to the long-lived shelving state before returning to the ground state [49]. This decay process of the $m_s = \pm 1$ states is not spin conserving. Meanwhile, electrons in the $m_s = 0$ excited state undergo a fast radiative transition back to the ground state which preserves the $m_s = 0$ population [50]. The population difference between the bright $m_s = 0$ state and the dark $m_s = \pm 1$ state can be resolved by time-resolved luminescence, shown in Figure 2b.

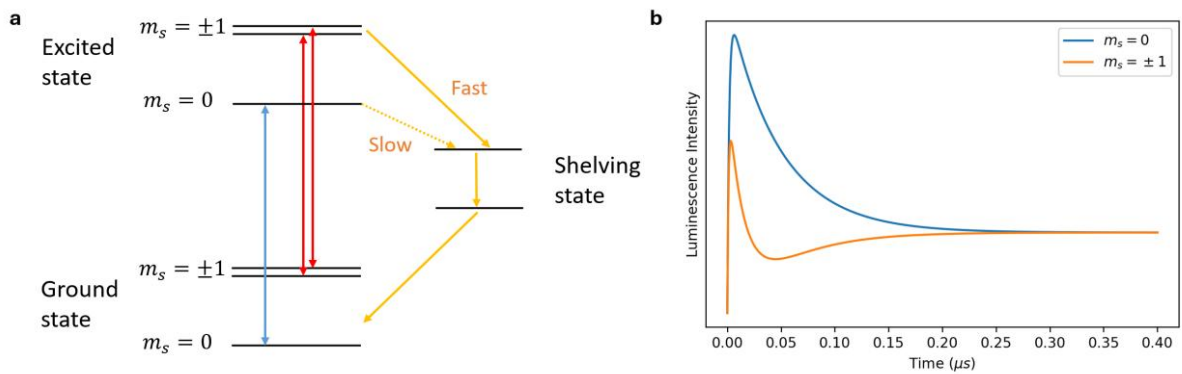

**Figure 2. a)** Energy level diagram of NV center consisting of ground, excited and shelving electronic states. Double-headed arrows between the ground and excited state show the spin-preserving transition. Solid and dashed arrows from excited to shelving state show the fast and slow transitions which allow for system initialization to the ground $m_s = 0$ state. **b)** Luminescence intensity against time elapsed when an excitation laser length of 0.4 $\mu s$ is turned on. The bigger area under the $m_s = 0$ curve shows that it is the "bright state". The intersystem crossing rate values used to simulate the plot are taken from [51].

When the nuclear spin is coupled to the electron spin, the electron energy level will be split into several nuclear sublevels due to the hyperfine coupling. This energy sublevel splitting only happens with the $m_s = \pm 1$ states due to non-zero hyperfine coupling. The presence of the nuclear sublevels can be observed as subpeaks in the optically detected magnetic resonance (ODMR) spectrum. Figure 3 shows typical ODMR spectra of NV centers when coupled with $^{15}$N and $^{14}$N separately [52]. Since $^{15}$N has spin quantum number of $I = 1/2$, we can observe the $2I + 1 = 2$ subpeaks in Figure 3a. The subpeak separation is measured at 3.1 MHz, which corresponds to the hyperfine coupling of $^{15}$N nuclear spins. $^{14}$N nuclear spins, with spin quantum number of $I = 1$, shows 3 subpeaks, each separated by 2.2 MHz, shown in Figure 3b. Observing these subpeaks allows for measurement of hyperfine coupling strength between the electron and nuclear spin, as well as determining the spin quantum number of the coupled nuclear spin.

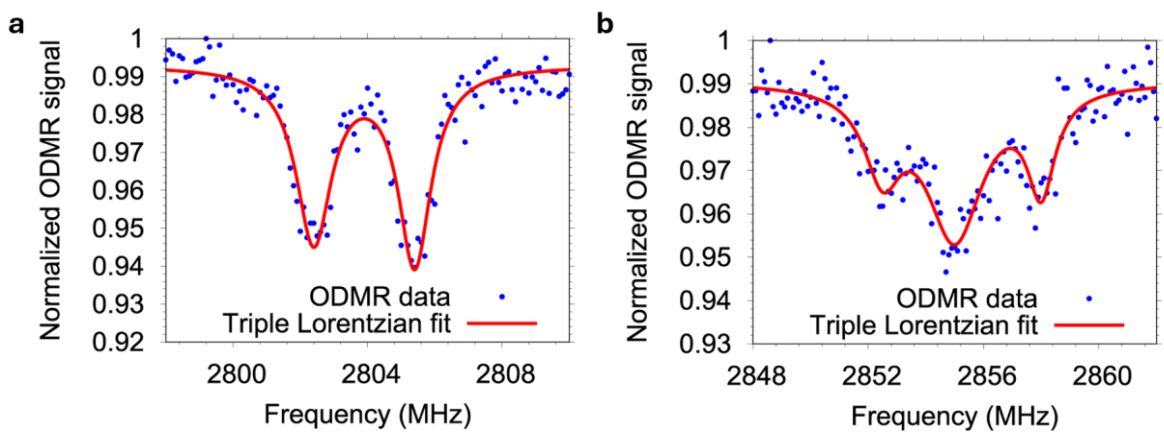

**Figure 3.** ODMR spectrum of NV centers coupled with a) $^{15}$N , b) $^{14}$N. 2 and 3 subpeaks are observed, corresponding to the respective spin quantum numbers of $I = \frac{1}{2}$ and $I = 1$. Reproduced with permission from

## 2.3. Nuclear-electron spin coupling and control

There are two types of nuclear-electron spin coupling in NV centers: intrinsic and extrinsic coupling. Intrinsic coupling is formed during the defect fabrication process [53, 54]. The electron spin defect is deterministically coupled with the implanted $^{14}$N nuclear spin and can exhibit large contact hyperfine interaction. This interaction strength may be resolved from the Optically Detected Magnetic Resonance (ODMR) spectrum if it is larger than the electron spin line width. The latter is limited by the inhomogeneous field electron spin coherence time, $T_2^*$. Extrinsic coupling with the $^{13}$C isotopes in the vicinity of the vacancy defect is determined by its small hyperfine coupling and usually couples the electron spin defect with the vicinity nuclear spins. Determining the magnitude of this hyperfine coupling can be achieved through methods such as Electron Spin Echo Envelope Modulation (ESEEM)-based hyperfine spectroscopy [55] and Carr–Purcell–Meiboom–Gill (CPMG) control sequence [56]. These methods enable the detection of weakly coupled electron-nuclear spin interactions by effectively filtering out unwanted background noise.

Nuclear spin control methods can be performed directly and indirectly. The direct resonant driving method can select spin states which requires substantial nuclear Zeeman splitting, high quadrupole interaction, and significant hyperfine interaction among nuclear sublevels. Spin states selection is achieved by driving the population transfer between electron and nuclear spins sublevels, facilitating usage of nuclear spin as memory qubit. Experimental evidence of direct resonant driving has been demonstrated with single shot quantum nondemolition measurements [47, 57, 58]. Successful implementation of direct resonant driving relies on properly setting the detuning parameter to isolate the coupled nuclear spin from the surrounding spin bath, thus shielding it from environmental interference.

Indirect nuclear spin control method is done through phase accumulation via electron spin manipulation. Due to the hyperfine coupling between electron and nuclear spin, nuclear spin precession and its phase information is conditioned on the spin state of the electron spin. This phase accumulation suffers from the problem of dephasing and dictated by $T_2^*$ time. Dynamical decoupling can be implemented to cancel the unwanted lattice spin bath interactions. In dynamical decoupling consists of train of pi-pulses with specified inter-pulse delay. Through rabi oscillation experiment, an appropriately chosen inter-pulse delay selectively decouples the spin bath, allowing the hybrid spin system to include multiple weakly coupled nuclear spin registers. Experimentally sweeping frequency leads to appropriate choice of frequency. When this specified frequency is selected, the electron spin will start to accumulate the phase of the selected nuclear spin [59,60], which leads to controlled nuclear spin rotation through electron spin manipulation.

Additionally, through smart interleaving of nuclear spin control, dynamical decoupling, correct tuning of phase and frequency of the radio frequency (RF) field, detection and control of weak electron-nuclear spin interactions can be achieved. In the next section, we will describe the basic mechanism by which electron-nuclear spin coupling can be manipulated.

## 3. Nuclear Spin Assisted Quantum Sensing Enhancement Protocol

### 3.1 Comparison between Nuclear spin assisted Protocol to Dynamical Decoupling

The motivation behind choosing nuclear spin assisted protocol over dynamical decoupling protocol is illustrated through the work of Wang et al. [61]. They implemented a simple nuclear spin assisted protocol using SWAP gate to map the NV state to the nuclear ancilla. Two SWAP gates form the delay window – coined delayed entanglement echo by the authors. This protocol works through application of specified RF frequency to generate entanglement between the target nuclear spin and the NV center (Fig 4a). The superiority of nuclear spin assisted protocol compared to dynamical decoupling (DD) protocol can be shown in Figure 4b. The dotted lines in Figure 4b denote the detected nuclear spin signals from both protocols. On select frequencies, such as the one at around 4968 kHz, standard DD control exhibit stronger sensed signal. However, on the $^{13}$C signal labeled C1, signal enhancement is observed. Additionally, $^{13}$C signals such as C2 and C3 are detected under nuclear spin assisted protocol and not under DD protocol, which are the nuclear spins that have zero perpendicular component of their hyperfine coupling ($A_j^\perp = 0$). The ability for nuclear spin assisted protocol to sense $^{13}$C signal with arbitrary value to its hyperfine coupling component is valuable in experiment motivated to determine the strength and the direction of the hyperfine coupling.

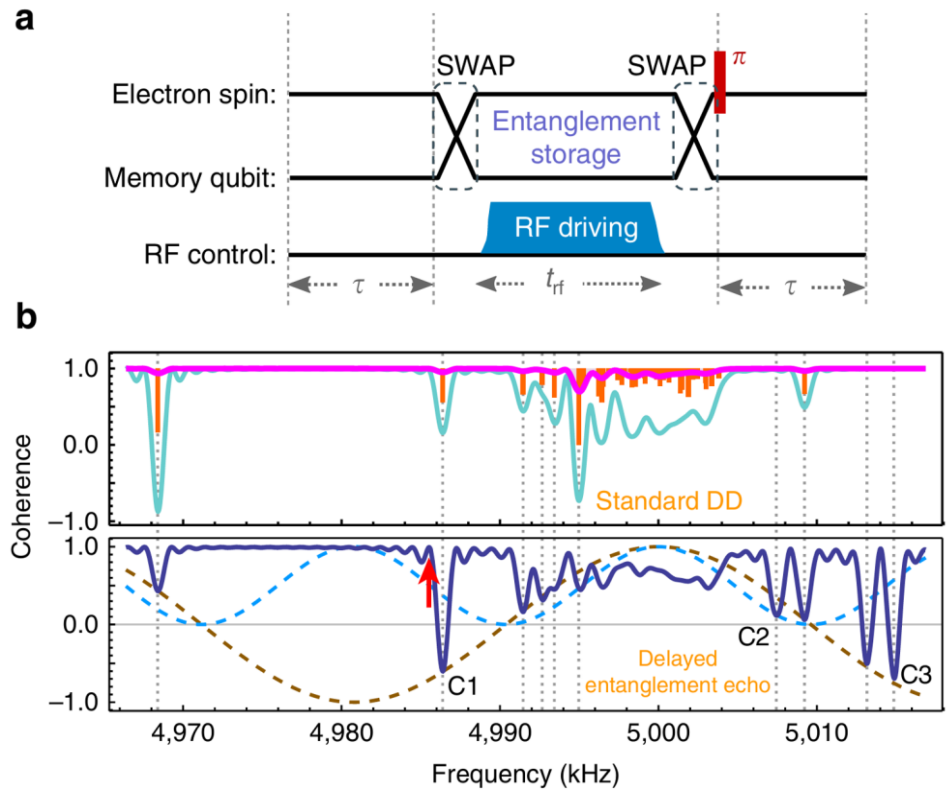

**Figure 4.** a) Schematic protocol of double SWAP gate to perform the entanglement information storage and retrieval. b) coherence signal comparison between standard dynamical decoupling (DD) methods against the delayed entanglement echo. C1, C2, C3 signals are stronger and C2 C3 signal that have $A_j^\perp = 0$ is resolved. Reproduced with permission from Wang, ZY., Casanova, J. & Plenio, M, Nat Commun 8, 14660 (2017); licensed under a Creative Commons Attribution (CC-BY) license [61].

## 3.2. The Foundation of Nuclear Spin Assisted Protocol

The fundamental mechanism of nuclear spin assisted quantum sensing enhancement lies in the high coherence of nuclear spins and its isolation from lattice spins, leading to improved electron spin coherence and enhanced quantum memory through nuclear spin qubit. The hybrid electron-nuclear spin control allows for the creation of a sharp frequency filtering system tuned specifically to isolate target nuclear spin from surrounding spins. As shown in Figure 5a, initialization of the NV sensor prepares the electron spin in the ground state, and the spin-locking technique transfers this polarization to the memory nuclear spin. Then the spatial resolution of the nuclear sensing is increased by the length of the time evolution under a gradient magnetic field, $t_g$. The tuning of the control scheme creates a filtering effect, which can be seen in Figure 5b. The filter linewidth, $\delta A$, and the filter bandwidth, $\Delta A$, are dependent on $t_g$ and sequence repetition, $F$. Increasing $t_g$ and $F$ will create a sharp filter for the detection of the electron-nuclear spin coupling. Increasing $t_g$ will increase the filter bandwidth, which allows the sequence to distinguish more distinct electron-nuclear spin coupling.

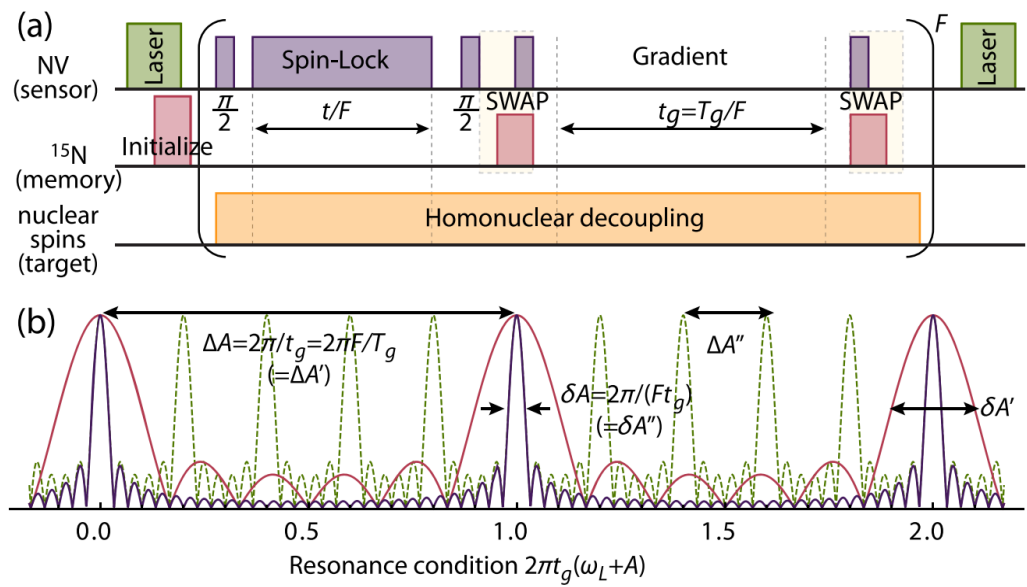

**Figure 5**. a) Control Sequence for memory transfer to 15N memory qubit. (b) The shape of the filter function is modified by adjusting different sequence parameters. Reproduced with permission from Ashok Ajoy et al, Phys. Rev. X **5**, 011001 (2015); licensed under a Creative Commons Attribution (CC-BY) 3.0 License [62].

Armed with this framework, Ajoy et al. [62] performed both 1D and 2D nuclear magnetic resonance (NMR) spectroscopy to showcase the basic control sequence and tenfold increase in spatial resolution of the sensed nuclear spins. The 1D sensing was done by sweeping the parameter filter time and the spin lock driving frequency. The 2D sensing was done by letting the target nuclear spin evolve during its diffusion time, transferring its polarisation back to NV electron spin which then can be detected subsequently. The nuclear spin sensing can be realized due to the diffusion of nuclear polarization to the surrounding nuclear spin through the homonuclear dipolar Hamiltonian. By analyzing the position and depth of the peaks in both the 1D and 2D spectra, the longitudinal and transverse nuclear-nuclear spin coupling can be retrieved, thereby resolving the polar coordinates of the target nuclear spin relative to the NV electron spin.

### 3.3. Filtering Function Optimization by Hyperfine Coupling

Knowing the value of hyperfine coupling ($A_j$) between the electron and nuclear spins optimizes the filtering function of the electron-nuclear spin control sequence without the need of rigorous guess and check attempts. The full measurement of the hyperfine tensor ($A_j$) can be obtained by the introduction of RF control field and RF decoupling field [49]. In a normal dynamical decoupling filtering sequence, the ability to address the target nuclear spin using its characteristic frequency may be compromised due to the presence of nuclear dipolar coupling. The application of RF control field helps decouple the target nuclear spin from its surrounding nuclear spins, thus enabling individual control of nuclear spins within coupled spin clusters. For instance, Wang et al. [63] extracted the values of various resonance frequencies that match the interrogated nuclear spins. They derived the magnitude of $A_j^{\parallel}$ and $A_j^{\perp}$ through the theoretical analysis and transformation of the electron spin basis from $m_s$ to $-m_s$. They also claimed that the direction of $A_j$ can be further determined by the application of an RF control field. When the direction of the applied RF field is in the same direction as the parallel component of the hyperfine tensor $A_j$ (denoted as $A_j^x$), the application of the RF control field does not affect the coherence of NV electron spin. This is because the RF control Hamiltonian commutes with the interaction of nuclear-electron spin Hamiltonian [64]. Figure 6a shows the coherence profile of the electron spin against the phase of the applied RF control field, $\phi_{rf}$. Since the RF phase controls both the effective direction of the RF field and the coherence of the electron spin, the correct RF phase is determined whenever the coherence is minimum resulting from turning the RF control on and off. The chosen RF phase concludes the direction of $A_j^x$. Figure 6b shows the relationship between the NV quantization axis ($B_z$), the Larmor precession of the nuclear spin and the components of the hyperfine tensor. Therefore, the hyperfine tensor $A_j$ is completely determined by the values of $A_j^{\parallel}$, $A_j^{\perp}$, and the direction of $A_j^x$.

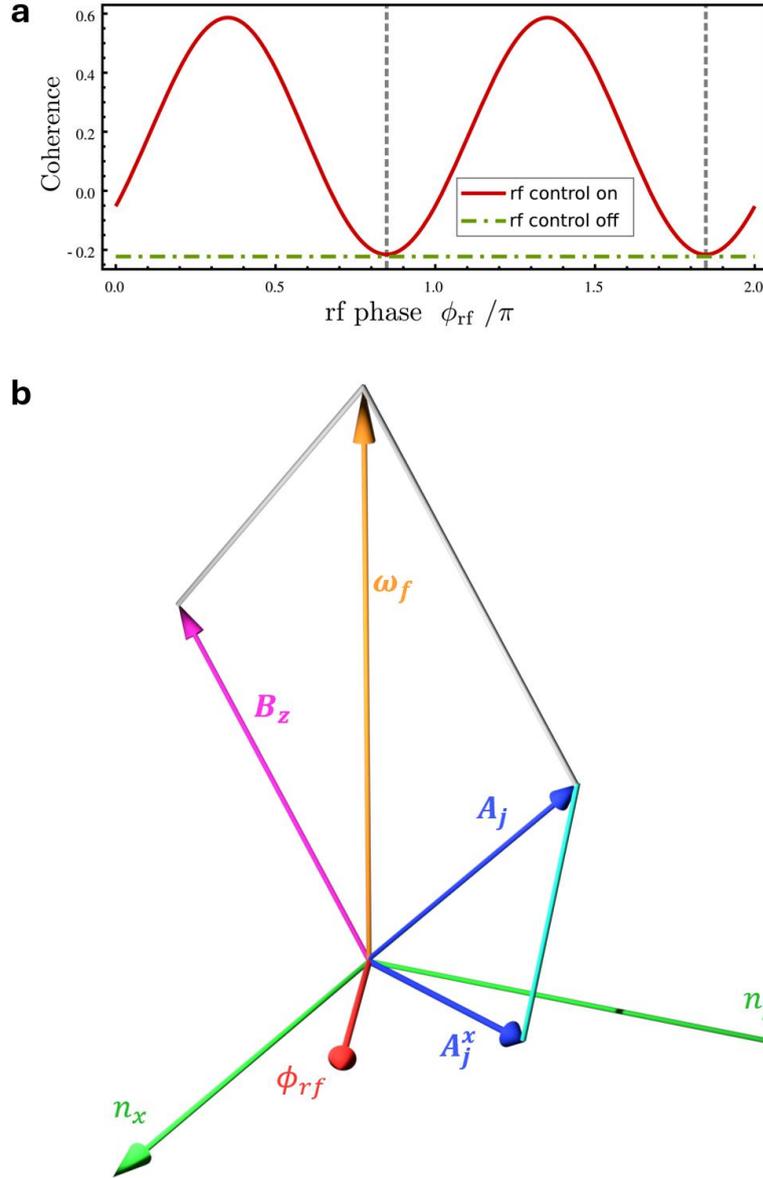

**Figure 6.** a) Coherence profile as a function of RF phase. When the coherence of RF control on curve coincides with RF control off curve, correct $\phi_{rf}$ is determined. Reproduced with permission from Zhen-Yu Wang et al, Phys. Rev. B **93**, 174104 (2016); Copyright 2016 American Physical Society [63]. b) Three-dimensional visualization of direction of $A_j^x$ and $A_j$. As $\phi_{rf}$ rotates on the $n_x$ and $n_y$ plane and its phase determined, the direction of $A_j^x$ is parallel to this direction. With the restriction imposed on $A_j^\parallel$, $A_j^\perp$, by their relationship with the applied magnetic field and the Larmor frequency, the hyperfine tensor $A_j$ is derived.

### 3.4. Nuclear Spin Assisted Protocol Beating Standard Quantum Limit

Tianyu et al showed the possibility of entangled hybrid spin system to beat the standard quantum limit (SQL) through projective measurements [65]. The initial state preparation of the entangled electron-nuclear spin state is important for the methodology and the result to exceed the SQL. Tianyu et al manages to initialize the system of NV center, $^{13}$C nuclear spin and $^{14}$N nuclear spin to a joint state of $|m_s = 0\rangle \otimes |m_{I,C} = -\frac{1}{2}\rangle \otimes |m_{I,N} = +1\rangle$ with 93.3% fidelity. The multi-step sequence of preparation of the NV$^-$ state to $m_s = 0$ and the polarization transfer of the nuclear spins with the electron spin between $^{13}$C nuclear spins and $^{14}$N nuclear spins are shown in Figure 7a, with the fidelity of each step to be 90.0%, 98.34% and 98.71% respectively. The prepared joint system acts as a Greenberger-Horne-Zeilinger (GHZ) state that is maximally metrologically useful and is used to prove that the entanglement-based interference pushes the system past the SQL limit. The quantum circuit involves CPhase and $C_n NOT_e$ gates, as shown in Figure 7b. The quantum Fisher information is given as $F_Q[\rho_{expt}] = Vis^2 \times N^2$, where $Vis$ is the interference visibility or amplitude of the interference pattern of the various one, two and tree spins entanglement, shown in Figure 7c. For all number of nuclear spins, Tianyu et al has achieved an experimental entangled interference result that resided between the SQL and Heisenberg limit (HL), shown in Figure 7d. The group also shows sub-SQL magnetic field sensitivity of two entangled nuclear spin in detecting an applied magnetic field. Similar interference visibility metric to the one used in Figure 7c was also used to measure the magnetic field variances. Through repeated measurement presented in Figure 7e, Tianyu et al has also shown sub-SQL magnetic field sensitivity in NV center utilizing electron-nuclear spin hybrid system.

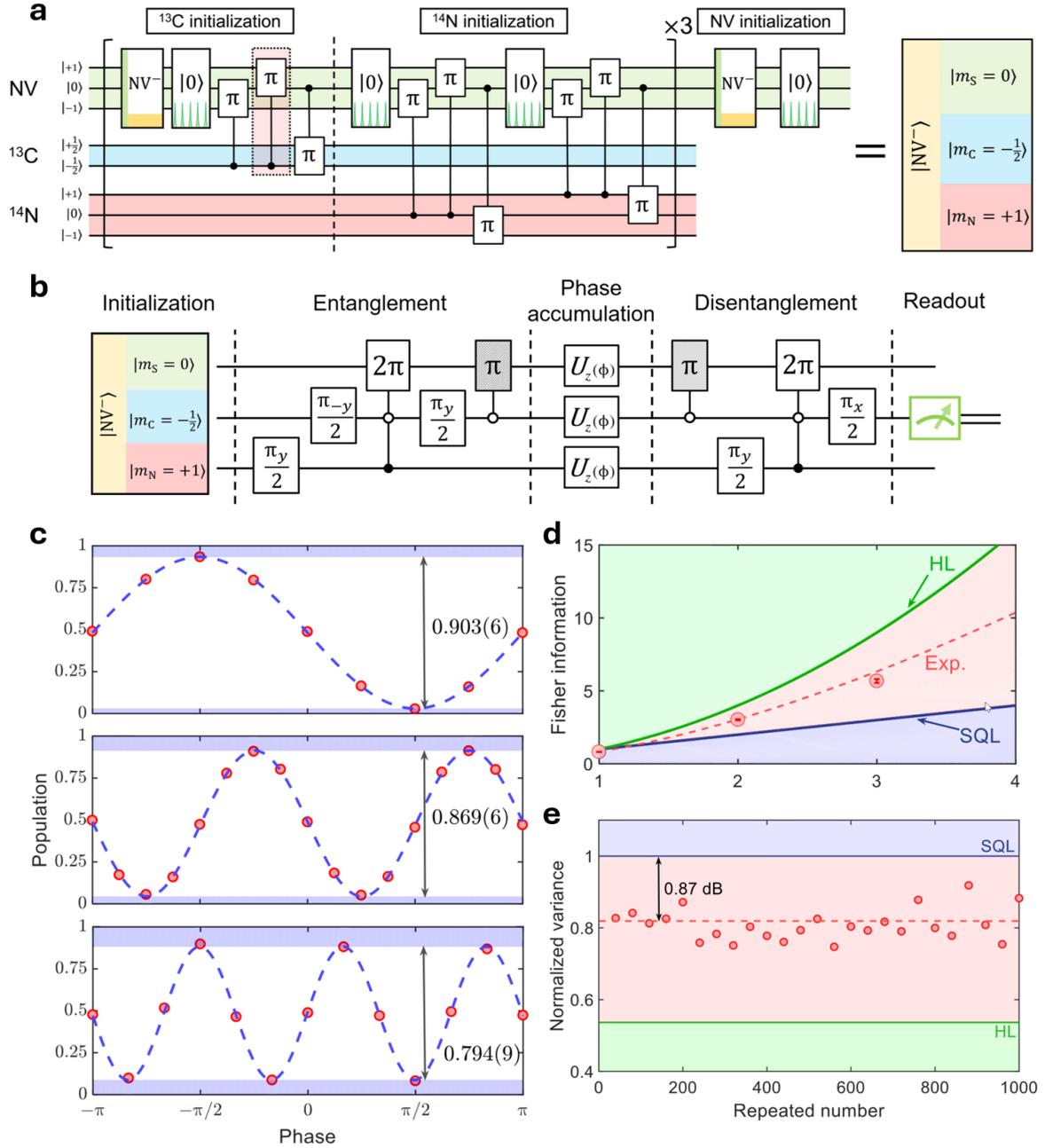

**Figure 7.** a) Overall sequence of state initialization of $|m_s = 0\rangle \otimes |m_{I,C} = -\frac{1}{2}\rangle \otimes |m_{I,N} = +1\rangle$. (b) Quantum circuit of a 1 electron and 2 nuclear spins interference. (c) Interference pattern of one, two and three nuclear spins from top to bottom. The interference visibility is given by the double-headed arrow. (d) Quantum Fisher information of the interferometer proposed by Tianyu et al group. Dashed line is the fitting of the metric against nuclear spin number, $N$, given by $N^2(0.91 \times 0.96^{(N-1)^2})$, compared to the SQL and HL limit of $N$ and $N^2$ respectively. (e) Magnetic field variance of the two-spin interference result. Dashed line is the experimental fit where the variances are normalised against SQL. Reproduced with permission from Tianyu Xie et al., Sci.

## 4. Applications of quantum sensing enhancement by nuclear spin-assist system

### 4.1. Nuclear Spin Spectroscopy

The hybrid electron-nuclear spin register can be implemented to perform high-resolution NMR spectroscopy to resolve target $^{13}$C nuclear spins. Pfender et al. [66] developed correlation spectroscopy by using the $^{14}$N nuclear spin as the memory qubit, enabling spectral resolution unrestricted by the limit of the electron spin time $T_1$. The uniqueness of this protocol lies in its utilization of the CROT gate rather than the SWAP gate. The benefits offered by the CROT gates encompass several aspects: the external B-field dependent quantum phase, the ability to address the final superposition state of the memory nuclear spin qubit with the acquired quantum phase, and the assurance that the electron spin state is in a certain spin projection. Figure 8 shows the performance of their technique by performing NMR spectroscopy of weakly coupled to their hybrid spin-register. By initializing the electron spin system to $|1\rangle$ instead of $|0\rangle$ state (Figure 8a), the measurement of the two target $^{13}$C spins can be resolved due to the shifting of RF pulse frequency by the strength of the hyperfine coupling strength. Increasing the sensing time and the memory storage time increased the signal contrast of the spectroscopy, as evidenced by Figure 8b.

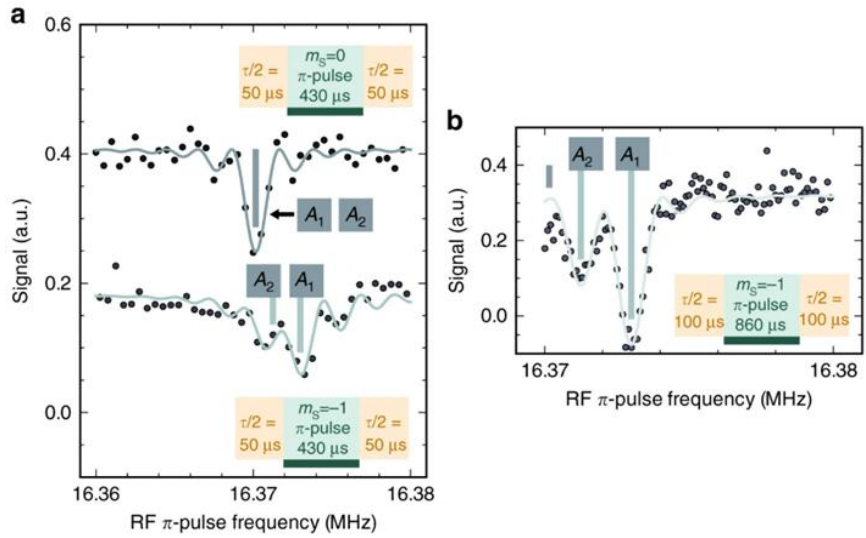

**Figure 8.** Spectroscopy of target $^{13}$C nuclear spins. **(a)** The two distinct hyperfine couplings are resolved when the electron spin state is at $m_s = -1$ state. **(b)** Signal contrast is improved when interrogation and $\pi$ pulse time is increased. Reproduced with permission from Matthias Pfender et al., Nat Commun 8, 834 (2017); licensed under a Creative Commons Attribution (CC-BY) license [66].

Zaiser et al. [67] demonstrated the ability of an entangled hybrid spin system to enable weak measurement of surrounding nuclear spins. In their paper, they developed tuneable sequences that allowed for full and zero entanglement between the NV spin and the $^{14}$N spin used as the memory qubit. In their protocol, CROT gate was similarly employed. The full protocol is shown in Figure 9a. The $\frac{\pi}{2}$ pulses prepare the nuclear spin in the superposition state, and the various $\pi$ pulses are the CROT gate that entangles the electron and nuclear spin to a certain degree or disentangles them fully. To show the effectiveness of their protocol, Zaiser et al. use stored phase information as a figure of merit and the comparison between zero and full entanglement degrees. Surprisingly, their protocol shows that the measurement of the decay constant against the duration of sensing time exhibits a certain degree of entanglement, which has little effect on the relaxation time of the electron spin, as shown in Figure 9b. Moreover, the phase information stored within the nuclear spin monotonously increases with the degree of entanglement, shown in Figure 9c, indicating an increased sensitivity of the entangled hybrid spin system in comparison with the single NV spin alone.

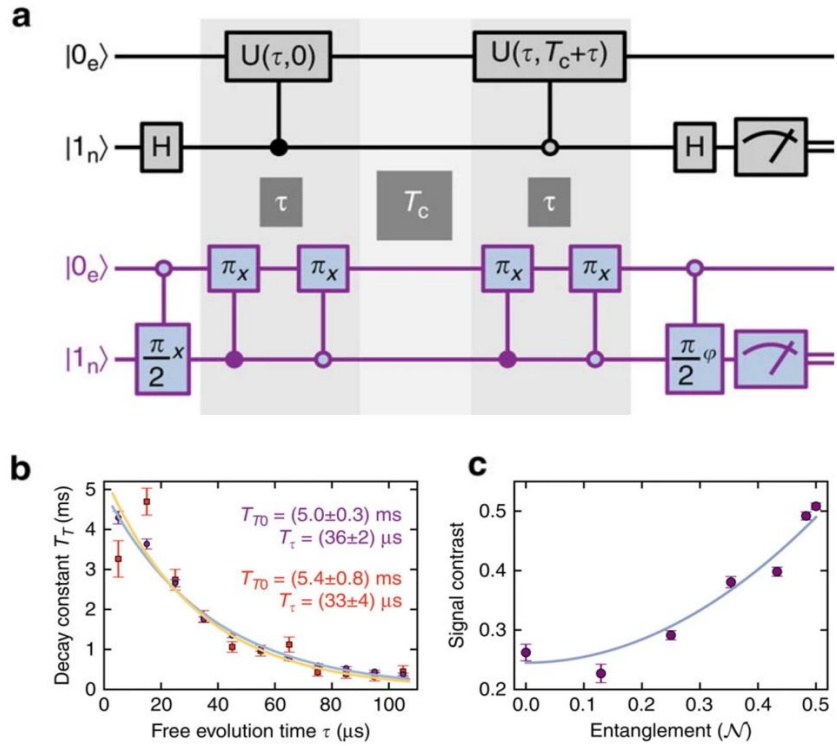

**Figure 9.** a) Quantum wire diagram and the corresponding $\pi$ and $\frac{\pi}{2}$ pulses form the CROT gates to control the entanglement between the electron and the nuclear spin. b) The orange and purple curves are the fit values of zero and full entanglement between electron and nuclear spin, which show negligible effect on the electron spin coherence. c) The monotonic increase of the signal contrasts as the entanglement between the electron and nuclear spin increases. Reproduced with permission from Sebastian Zaiser et al., Nat Commun 7, 12279 (2016); licensed under a Creative Commons Attribution (CC-BY) license [67].

Inspired by the work of Pfender et al and Zaiser et al, Chen et al utilized the intrinsic $^{14}$N nuclear spin of NV center as a memory to store the coherence of the electron spin [68]. By protecting it from both thermal and quantum noise of the surrounding $^{13}$C nuclear spins bath, the dephasing noise $T_2^*$ can be extended. The pulse sequence to suppress both thermal and quantum noise can be understood from Figure 10a. The green box in Figure 10a represents the 1$^{st}$ half of the transfer operation to store the coherence state of the electron spin into the intrinsic $^{14}$N nuclear spin. The 2$^{nd}$ half of the transfer operation is the single conditional gate C-R$_x$ gate after the Free evolution blue box to reduce the experimental noise during the transfer operation. The hard pulses implementation of the quantum circuit highlighted in green box is shown in Figure 10b. The suppression of quantum noise during phase accumulation of the free evolution highlighted in the blue box is achieved by dividing the evolution to N repetitions, with an $R_z^\pi$ gate inbetween, shown in Figure 10c. The implementation of this $R_z^\pi$ gate can be realized through simple DD sequence such as XY8 pulse sequence. Figure 10d shows the signal of the experimental results done at magnetic field of 10 $mT$. The extracted $T_2^*$ values of the Ramsey, decoupled thermal noise, and decoupled thermal and quantum noise to have the values of 1.8, 4.2, and 7.8 $\mu s$ respectively, showing substantial increase in electron dephasing time by utilizing the $^{14}$N nuclear spin as memory register.

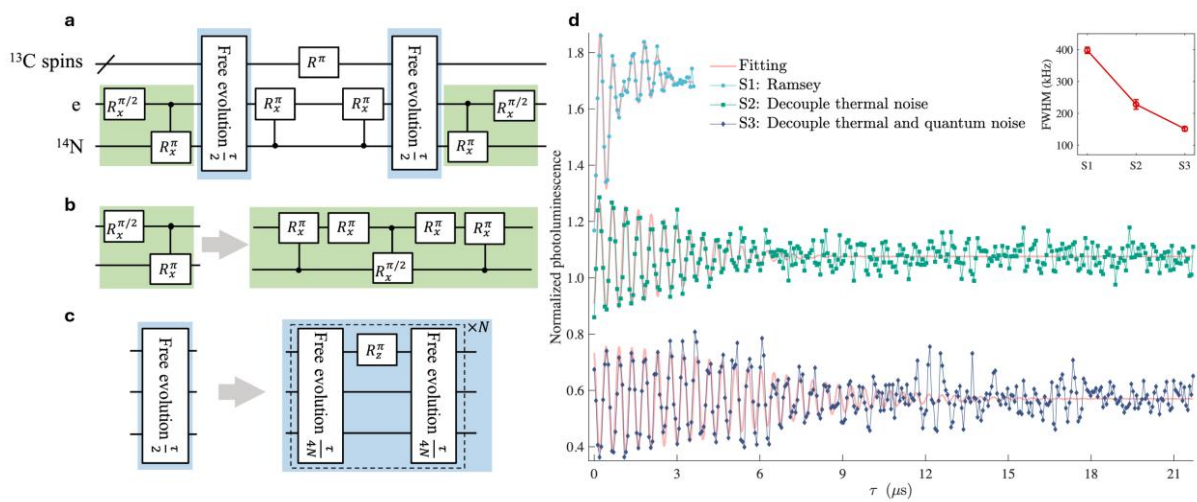

**Figure 10**. a) Quantum circuit showing the thermal and quantum noise suppression protocol. b) Transfer operation, highlighted in green, is realized by hard pulses. c) Inserting $R_z^\pi$ gate inbetween two blocks of free evolutions and repeats itself N times to suppress quantum noise. d) Normalized photoluminescence signals of the three methodologies; ordinary Ramsey protocol, decoupled thermal noise protocol, and decoupled thermal and quantum noise protocols. Red lines are fitted to the fitting function. Inset shows the Full Width Half Maximum (FWHM) of each protocol. Reproduced with permission of Xin-Yu Chen et al., Phys. Rev. B 108, 174111 (2023). Copyright 2023 American Physical Society [68].

**4.2 Atomic Imaging Sensing**

Abobeih et al. [69] pushed the concept of nuclear spin detection using the hybrid spin register to perform a 3D mapping of 27 nuclear spin clusters at Angstrom resolution. The basic concept revolves around using an electron spin NV sensor to detect multiple nuclear-nuclear spin couplings. The concept of the experiment is explained in Figure 11. Instead of a fixed electron-nuclear spin coupling as the basis for the hybrid spin register, Abobeih et al. detected 1 out of the 27 nuclear spins at a time as the memory qubit of the hybrid register at frequency RF1. This nuclear spin acts as a probe that scans the other 26 nuclear spins using the double resonance technique, acquiring the frequency RF2 and bypassing the short coherence time of the electron spin sensor. By varying the interrogation time $t$, frequency RF1 and RF2, 3D spectroscopy can be performed, and 3D data set that consists of the connections of various nuclear spins can be obtained.

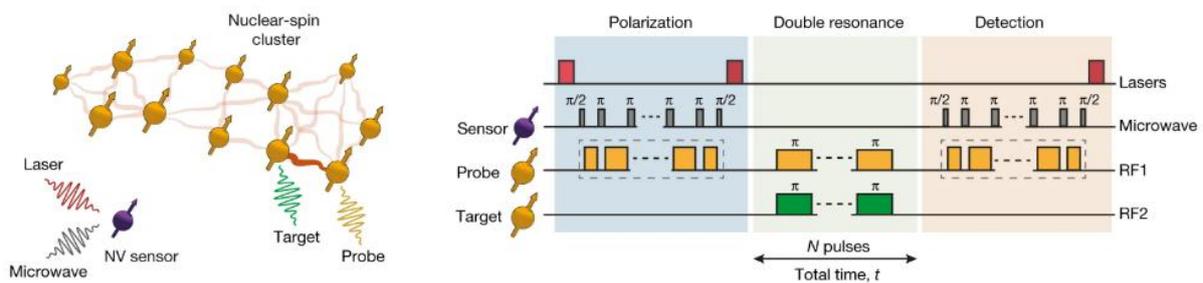

**Figure 11.** Schematic Diagram of NV sensor coupled to 1 nuclear spin in the nuclear spin control as the target to probe the rest of the nuclear spin and the sequence control. Reproduced with permission from MH Abobeih et al., Nature 576, 411–415 (2019). Copyright 2019 Springer Nature [69].

The further study of imaging clusters of spins is discussed by Stolpe et al. [70]. They developed a correlated sensing sequence and treated the nuclear spin clusters as a graph network, measuring their network characteristics and their characteristic spin frequencies. The graph network consists of fifty $^{13}$C nuclear spins. Their protocol, as seen in Figure 12, showcases the spin-chain sensing of 5 nuclear spins linearly connected to the electron spin sensor. Through the concatenated spin-echo double resonance (SEDOR) method, they recover the individual coupling between nuclear spin pairs by continuously extending the chain and mapping back the signal through the spin chain to maintain the signal information. Extending this concept to 50 nuclear spin networks, Stolpe et al. successfully performed 3D imaging of the nuclear spin network. However, the protocol requires a cryogenic temperature to achieve its increased signal sensitivity.

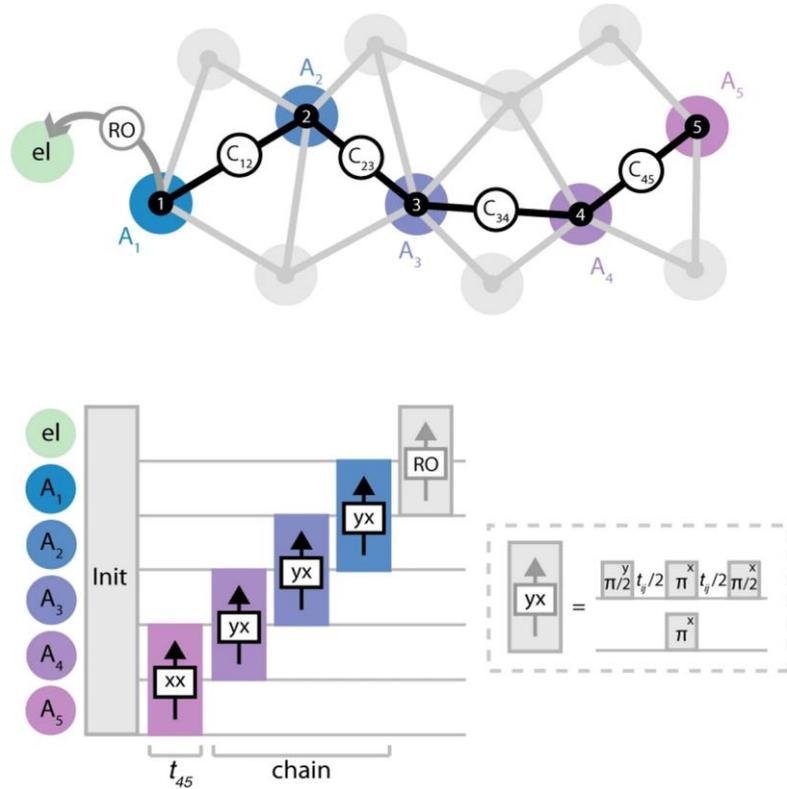

**Figure 12.** The graph network diagram showing 5 $^{13}$C nuclear spins connected to the main electron spin as a spin-chain and the brief cascading SEDOR protocol to send the signal back to for electron spin readout. Reproduced with permission from GL Van de Stolpe et al., Nat Commun 15, 2006 (2024); licensed under a Creative Commons Attribution (CC-BY) license [70].

## 4.3 Magnetic Field Sensing

Another implementation of the hybrid electron-nuclear spin register is to increase the spectral resolution of arbitrary magnetic field sensing [71]. Strategic incorporation of CNOT gates in the pulse sequence, alongside single and double gate readout retrieval, can enhance contrast and boost experiment readout efficiency as well as spectral resolution. Rosskopf et al. performed the detection of arbitrary spectroscopy of AC signals without and with the memory qubit (Figure 13a and 13b respectively). The timescale of the signal decay in Figure 13a was measured to be $\sim 0.5\ ms$ which is bounded by the electron spin time $T_1$. Conversely, when assisted by the nuclear spin, the non-decaying oscillation continued more than 30 ms as shown in Figure 13b, surpassing the limitation of the nuclear spin time $T_1$ by two orders of magnitude. Figure 13c is the Fourier spectrum of the signal in Figure 13a, where the inset is the spectrum of the signal in Figure 13b. It can be observed that the linewidth has improved by a factor of 100 when nuclear spin is involved as the memory qubit.

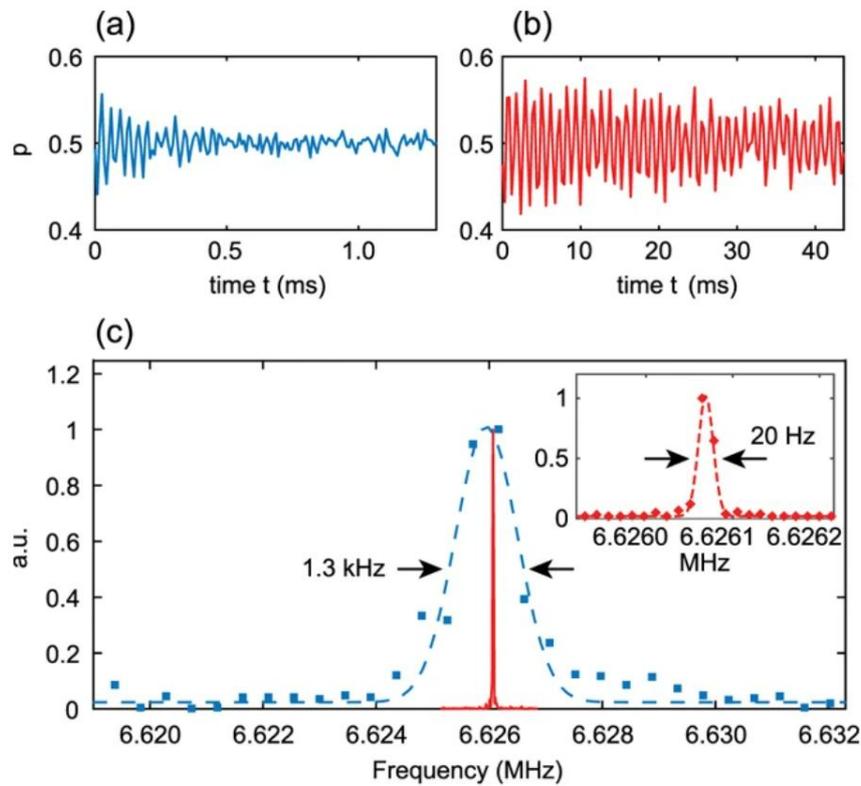

**Figure 13**. Spectroscopy of an external AC field signal. **(a)** Without a nuclear memory qubit. **(b)** with a nuclear memory qubit. **(c)** Fourier transform of plots (a) and (b), showing improved linewidth and resolution of the detected AC field. The inset is the zoomed-in version of the red curve in Figure C. Reproduced with permission from Tobias Rosskopf et al., npj Quantum Inf 3, 33 (2017); licensed under a Creative Commons Attribution (CC-BY) license [71].

The sensitivity of magnetic field detection can be increased by using the accumulated phase information from the nuclear spin. Coto et al. [72] proposed a simple double CROT gate to map the signal to and from the electron and nuclear spin, as shown in Figure 14a. During the sensing step, the sensing time $\tau$ governs the evolution of the nuclear spin signal $\langle I_z \rangle$. Figure 14b shows the sensitivity of the nuclear spin signal against a small change in interrogation time at around $\tau = 2.0\ \mu s$. Comparing their technique (black curve) against the standard Ramsey spectroscopy (blue curve) in Figure 14c, the standard deviation of detected magnetic field $\Delta B$ is improved in certain regimes of the interrogation time, showing its improved sensitivity in magnetic field measurements.

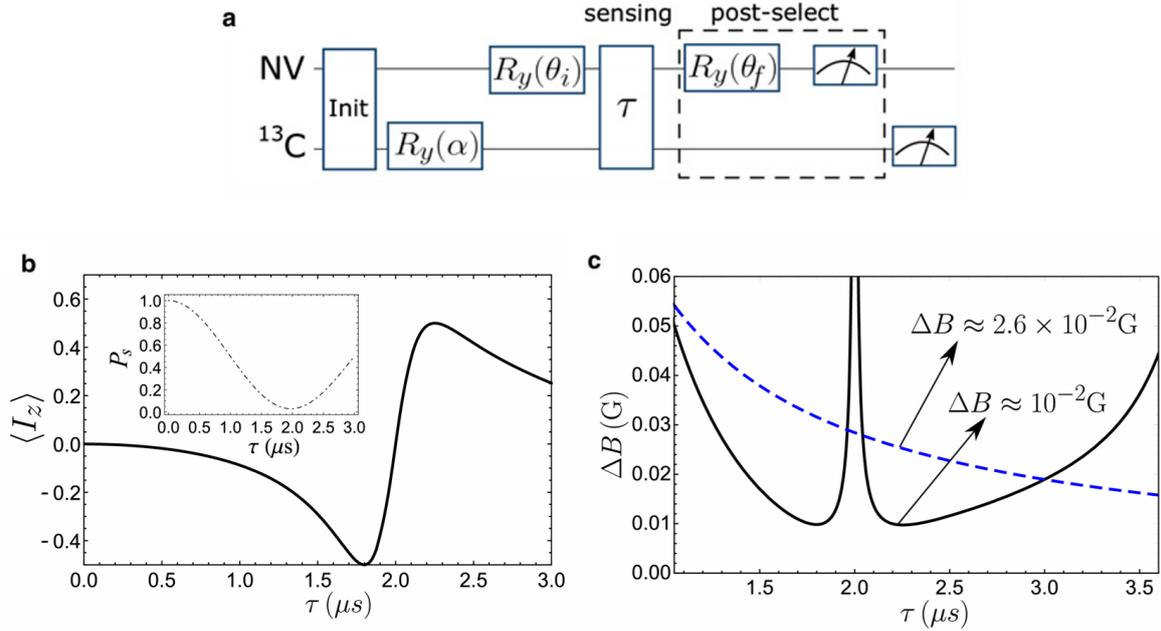

**Figure 14.** a) Proposed protocol consisting of electron-nuclear entanglement, sensing, and readout protocol. b) Variation of the nuclear spin information, against the sensing time $\tau$. c) the standard deviation of detected magnetic field sensing against sensing time $\tau$. The black and blue curve denote the results of the proposed protocol and the Ramsey method respectively. Reproduced with permission from Raúl Coto et al., Quantum Sci. Technol. 6 035011 (2021). Copyright 2021 IOP Publishing Ltd [72]

Qiu et al. [73] explored the behavior of the hyperfine interaction between the electron spin sensor and the nuclear spin qubit. Noticing that the entanglement created between the electron-nuclear spin pair is sensitive to the change in the transverse magnetic field, it serves as the basis for the magnetic field angle sensing approach. As the transverse magnetic field approaches the perpendicular direction, the hyperfine splitting approaches 0 almost linearly. This allows for the detection of the nuclear spin sublevel due to the electron spin echo envelope modulation (ESEEM) effect, which can be easily detected by using a typical spin echo sequence. The result of the angle sensitivity experiment can be seen in Figure 15. Figures 15a and 15b show the experimental and theoretical results of the spin echo signal when the sequence time $\tau$ and angle of the transverse magnetic field $\theta_B$ are swept. Figure 15c shows the sensitivity of the protocol to the change in transverse magnetic field. Qiu et al. also performed numerical computations to compare the performance of their protocol against conventional methods. Increasing the sequence integration time $\tau$ expands the regime of the protocol for magnetic angle sensing (Figure 15d). Both methods are complementary to each other, where the proposed protocol works best near a 90-degree angle, while the conventional method works best at angles away from 90-degree.

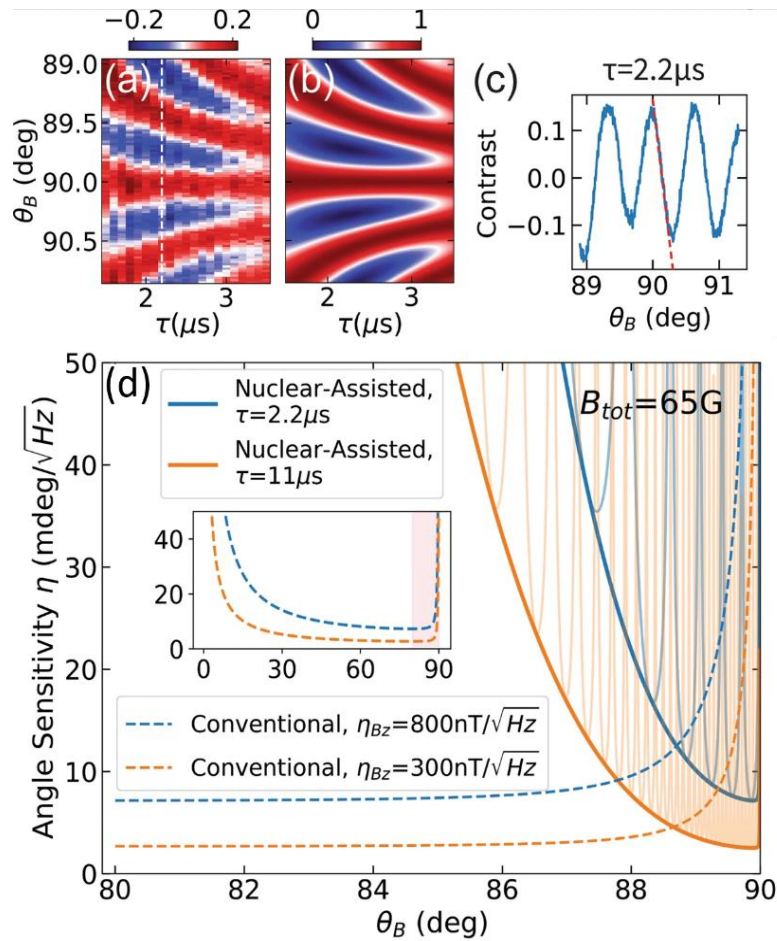

**Figure 15**. a) Experimental and b) simulation results of the spin echo signal. **(c)** Line-cut of (a) to show angle sensitivity (d) Sensitivity comparison between nuclear spin assisted and conventional protocols. Reproduced with permission from Ziwei Qiu et al., npj Quantum Inf 7, 39 (2021); licensed under a Creative Commons Attribution (CC-BY) license [73].

**4.4 Nuclear Spin Gyroscope**

Beyond lab-scale progress, the scope of potential applications of nuclear spin sensitivity enhancement toward commercialization has also been extended. The argument for NV center to become a competing rotation sensor compared to a mass-produced microelectromechanical system (MEMS) gyroscope has been ongoing [74]. The problem of miniaturization, as well as enhancing its sensitivity and long-term stability, has been the limiting factor. To compete with the current use of MEMS gyroscope, Soshenko et al. [75] and Jarmola et al. [76] independently demonstrated the proof of principle gyroscope utilizing nuclear spin in NV center. Using the double quantum nuclear Ramsey measurement, the precession of $^{14}$N nuclear spin is monitored and plotted, as shown in Figure 16a. A preselected delay time is chosen as the reference point, and any fluctuations relative to the measured signal because of performing rotation on the sensor result in a gyroscopic reading. Despite the NV center gyro sensor' signal noises, Soshenko et al shows its signal output is comparable to MEMS gyro sensor output, as shown in Figure 16b, demonstrating its promise as an inertia sensor through the utilization of nuclear spin in NV centers. Similarly, Jarmola et al reproduces the result of nuclear spin gyroscope by showing the linear fit of the nuclear spin gyroscope fluorescence against rotation rate of the gyroscope platform and comparing the rotation rate signal of the nuclear spin gyroscope gathers and the table, as shown in Figure 16c and 16d.

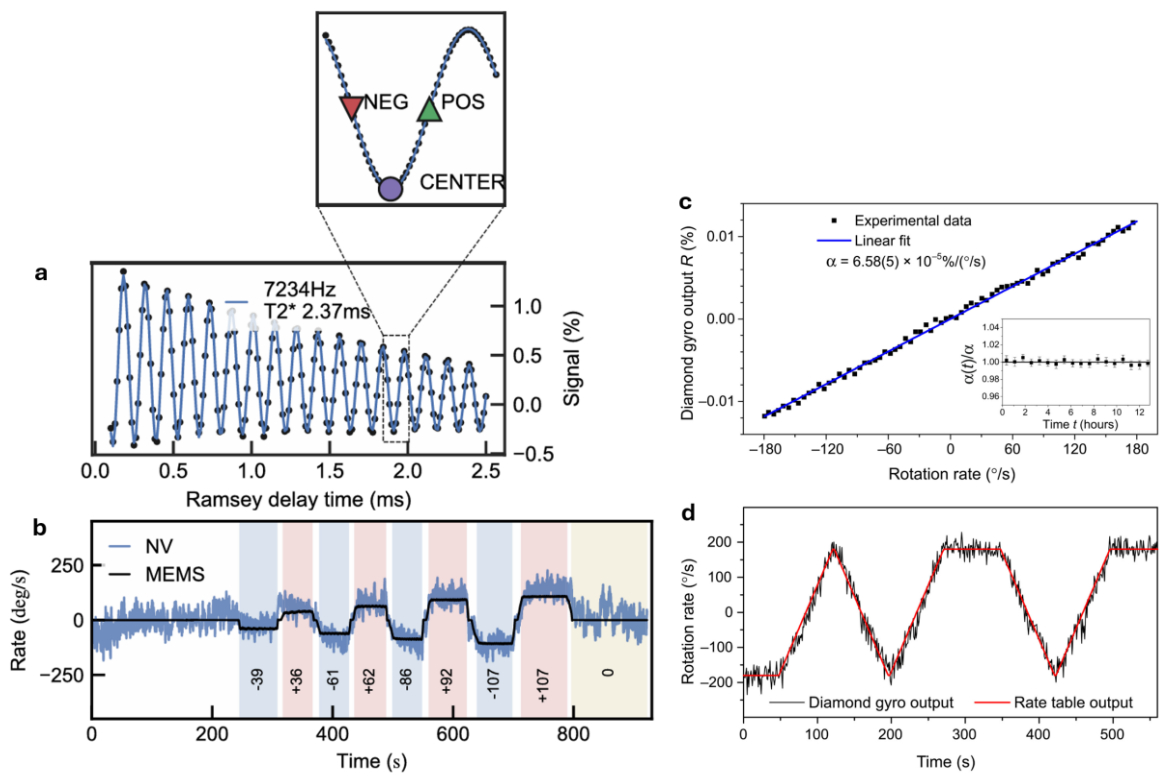

**Figure 16**. Nuclear spin gyroscope result. a) Ramsey measurement result that shows the nuclear spin coherence time. Inset shows the working point and the indication of positive and negative gyro sensor output relative to the signal of the center point. Reproduced with permission from Vladimir V Soshenko et al., Phys. Rev. Lett. **126**, 197702 (2021). Copyright 2021 American Physical Society [75]. b) Gyroscopic result of NV sensor (blue line)

and MEMS sensor (black line) against experiment time. The shaded area shows the rotation rate reading from MEMS sensor. c) Fluorescence signal measured against the rotation rate of the gyroscope platform. d) Rotation rate from the diamond nuclear spin gyroscope and the table output is plotted together, showing agreement with each other. Reproduced with permission from Andery Jarmola et al., Sci. Adv.7,eabl3840 (2021); licensed under a Creative Commons Attribution- NonCommercial (CC-BY-NC) 4.0 license [76].

## 5. Closing Statements and Outlook

Sensitivity enhancement on quantum sensors based on solid-state spin has been advancing at a steady pace. The early advancement in incorporating nuclear spin improves the performance of electron spin as a quantum sensor by extending the memory lifetime and coherence time. The early challenges of nuclear spin incorporation into the electron spin to form the hybrid spin register come from identifying the hyperfine interaction. Establishing the performance of the hybrid spin register comes from the improvement of the sequence protocol the nuclear $T_2$ times to nuclear $T_1$ times. Different kinds of gate operations are explored to create a distinct decoupling protocol that enhances sensing, ultimately improving its sensitivity compared to its bare electron spin sensor counterpart. With a better understanding of nuclear spin control, nuclear spin performance as memory qubits in these hybrid registers allows for higher spectral resolution in atomic imaging and auxiliary magnetic field sensing.

Moreover, NV centers as the platform for quantum sensing enhancement provide a good premise for other solid-state systems to shine. The most recent development of 2D materials such as hexagonal Boron Nitride (hBN) shows promising coherent electron spin control at room temperature [77]. ODMR spectrum of electron-nuclear spin coupling in hBN has been resolved [78], and coherent nuclear spin control paves a promising future for enhancing hBN's quantum sensing capability [79]. It is conceivable that leveraging the nuclear spins in hBN for quantum sensing enhancement could follow the same footprint as the development of NV centers.

Ultimately, the increase in NV center sensitivity through the nuclear spin register paves the way for commercialization. The increased sensitivity of magnetic field sensing may lead to the proper development of inertial navigation sensing [80], and apt for military uses [81]. High-resolution RF sensing [82] can also be achieved and become the catalyst for developing quantum radio-frequency analyzers [83]. The challenge of commercializing nuclear spin sensing technology lies in its usability and scalability. Most of the advances discussed in this review are still done in laboratory settings. For scalable applications, it's essential to scale the control of nuclear spin as part of an NV ensemble rather than a single spin to attain sensitivity enhancement comparable to conventional nuclear magnetic resonance outcomes. For instance, the control of bulk $^{13}$C as part of the nuclear spin sensing enhancement strategy has been actively investigated [84,85]. A certain degree of control has been attained through the atomic imaging of 27 and 50 nuclear spin clusters and the showcase of sensitivity has been displayed through transverse magnetic field sensing. As nuclear spin control is pivotal in pushing the frontier in nuclear spin-assisted quantum sensing, future research endeavors may lie in the ability to achieve multiple nuclear spin control through one single electron NV sensor, thus unlocking bigger, previously inaccessible spectral regions that allow for increased sensing sensitivity, as showcased by Stolpe et al. [70].

## 6. Acknowledgments

This work was supported by Singapore National Research foundation through QEP Grants (NRF2021-QEP2-01-P01, NRF2021-QEP2-01-P02, NRF2021-QEP2-03-P01, NRF2021-QEP2-03-P10, NRF2022-QEP2-03-P11), ASTAR IRG (M21K2c0116), MOE grant (MOE-T2EP50221-0005 and MOE-T2EP50222-0018).